\documentclass[english,floatfix,twocolumn,showkeys,superscriptaddress,nofootinbib,aps,prd]{revtex4}
\usepackage[utf8]{inputenc}
\usepackage[T1]{fontenc}
\usepackage{lmodern}
\usepackage{amsmath}
\usepackage{amssymb}
\usepackage{graphicx}
\usepackage{esint}
\usepackage{longtable}
\usepackage{dcolumn}
\usepackage{babel} 
\usepackage{csquotes} 
\usepackage{color} 
\usepackage{hyperref}
\usepackage{setspace}
\begin{document}

\title{Shadows of black holes at cosmological distances in the co-varying physical couplings framework}

\author{R. R. Cuzinatto}
\email{rodrigo.cuzinatto@unifal-mg.edu.br}
\affiliation{Instituto de Ciência e Tecnologia, Universidade Federal de Alfenas, \\ Rodovia José Aurélio Vilela,
11999, CEP 37715-400 Poços de Caldas, MG, Brazil}

\author{C. A. M. de Melo}
\email{cassius.melo@unifal-mg.edu.br}
\affiliation{Instituto de Ciência e Tecnologia, Universidade Federal de Alfenas, \\ Rodovia José Aurélio Vilela,
11999, CEP 37715-400 Poços de Caldas, MG, Brazil}

\author{Juliano C. S. Neves}
\email{juliano.c.s.neves@gmail.com}
\affiliation{Instituto de Ciência e Tecnologia, Universidade Federal de Alfenas, \\ Rodovia José Aurélio Vilela,
11999, CEP 37715-400 Poços de Caldas, MG, Brazil}


\begin{abstract}
The co-varying physical couplings (CPC) framework states that physical parameters like  
the speed of light in vacuum $c$, the Newtonian constant $G$, and the cosmological constant $\Lambda$ 
could indeed vary with the spacetime coordinates $x^{\mu}$.
Here, we assume a temporal variation, that is, $c(t),G(t)$ and $\Lambda(t)$.
We show that the McVittie spacetime, a black hole in an expanding universe, is a solution of the CPC framework
providing naturally an important parameter of the model.
Then, we calculate the shadow angular radius of this black hole at cosmological distances.
A black hole shadow in the CPC context could be either larger or smaller than the same shadow in the
standard cosmology. It depends on how the set $\{ c,G,\Lambda \}$ varies with time or with the cosmic expansion.
\end{abstract}


\maketitle


\section{Introduction} \label{sec:Intro}

Cosmology hints that modifications to the general relativity (GR) 
description of gravitational interaction should be implemented. 
In fact, a plethora of observational windows indicate that dark energy should dominate the energy budget 
of the universe today. From SNe Ia data \cite{SupernovaSearchTeam:1998fmf,SupernovaCosmologyProject:1998vns}, 
to cosmic chronometers $H(z)$ data \cite{Magana:2017nfs}, to BAO data \cite{SDSS:2005xqv}, 
to fraction of gas in galaxy cluster data \cite{SPT:2021vsu}, to GRBs \cite{Liu:2022inf}, to Quasars \cite{Lusso:2020pdb}, 
to even CMB data \cite{Planck:2018vyg}, all these observational sets indicate that a cosmological constant plays 
the role of dark energy leading to the accelerated expansion of the recent universe. However, 
there is conflict between the value of the cosmological constant $\Lambda$ as estimated from the late-time 
cosmology (SNe Ia data) and from the early universe (given by the CMB power spectrum). 
This is the Hubble tension \cite{DiValentino:2021izs} that adds to other problems of standard cosmology and GR such as the $S_8$ problem \cite{Abdalla:2022yfr} and discrepancies in the dark matter distribution \cite{Klypin:1999uc}. On top of that, the very nature of cosmological constant is an open problem, which constitutes an embarrassment from the theoretical point of view \cite{Carroll:2000fy}. 

The early universe also presents some challenges for the description offered by GR. 
Even though the standard model or the hot Big Bang scenario does successfully predict the CMB
 \cite{Planck:2018vyg}, the primordial nucleosynthesis of the light elements \cite{Coc:2017pxv} 
 and the Hubble flow \cite{SupernovaSearchTeam:1998fmf,SupernovaCosmologyProject:1998vns}, 
a solution for the early universe puzzles (including the horizon problem, the flatness problem and lack of predictions regarding structure formation) asks for a mechanism like the inflationary framework \cite{Starobinsky:1980te,Guth:1980zm,Linde:1981mu,Mukhanov:1981xt} (or alternatives like the ekpyrotic 
cosmology \cite{Lehners:2008vx}). This add-on framework to the standard model is not free from its limitations. The details of the particular model driving inflation is yet to be laid down. Even the Starobinsky inflationary model \cite{Starobinsky:1980te}, favored by Planck data \cite{Planck:2018jri}, admits some room for improvement since there seems to be indications of a running of the spectral tilt. 

Given these challenges faced by the standard cosmological from GR, it is argued that we are in the need 
of an improved description of gravitational interaction; one that encompasses GR and its successes while 
also addressing its problems (such as the nature of dark energy). This is the goal of the extended description of gravitation dubbed modified gravity.

Modifications of gravity were proposed very early after the inception of general relativity; 
the Kaluza-Klein model is an example \cite{Kaluza:1921tu,Klein:1926tv}. 
Extensions of the underlying spacetime manifold are also a family of modified gravity 
proposals \cite{CANTATA:2021ktz}. Among this class we single out Riemann-Cartan models that naturally 
accommodate the description of spin \cite{Minkevich:2016wya}; the teleparallel 
equivalent of GR \cite{Aldrovandi:2013wha}, and Lyra scalar-tensor gravity \cite{Cuzinatto:2021ttc}. 
A popular modified gravity scenario is the $f(R)$ theories (see e.g. Refs.~\cite{Clifton:2011jh,Sotiriou:2008rp,Nojiri:2010wj,Capozziello:2011et} for reviews); they extend the Einstein-Hilbert action to include other invariants built from the curvature scalar $R$. Theories including higher-order derivatives of $R$ were also proposed based on the reasoning that 
an effective description of gravity is required in the ultraviolet regime, where singularities appear 
and a quantum version of GR is in order (see Ref. \cite{Buchbinder:2021wzv} and references therein). 
The so called $f(R,\partial R,\dots,\partial ^n R)$ gravity was explored in a number 
of papers \cite{Cuzinatto:2016ehv,Cuzinatto:2018chu} and, in particular, Ref. \cite{Cuzinatto:2016ehv} 
shows that $f(R,\partial R,\dots,\partial ^n R)$ theories are equivalent to a particular 
class of scalar-tensor theories \cite{Faraoni:2004pi}. The latter are also an alternative to GR; 
they consider a scalar field $\phi$ (or multiple of them) alongside the 
connection $\Gamma^{\sigma}_{\mu\nu}$ as degrees of freedom geometrizing the 
gravitational interaction. The paradigm of scalar-tensor theories is the Brans-Dicke theory \cite{Brans:1961sx}.

Brans and Dicke assumed that the scalar field $\phi$ is the reciprocal of an effective gravitational 
constant $G$ non-minimally coupled to $R$. Their attempt was to fully implement 
Mach's principle \cite{Barbour:2010dp} in general relativity. 
Brans-Dicke model brings about the possibility of the gravitational coupling being a spacetime function 
instead of a genuine constant. Other couplings appearing in 
Einstein-Hilbert action---besides $G$---are the cosmological constant $\Lambda$ and the speed of light $c$. 
Once one relax the constancy of $G$, it is also natural to ask oneself about the 
possibility of $\Lambda$ and $c$ being functions of the spacetime coordinates. 
It is perhaps easier to accept a varying $\Lambda$---its nature escapes  us after all, and it might 
be a slowly varying quintessence field as well \cite{Caldwell:1997ii}---; not so much as for the case of $c$. 
Ellis and Uzan \cite{Ellis:2003pw} are particularly skeptical of varying speed of light (VSL) scenarios. The mentioned
authors emphasize that proposers of VSL models should minimally be careful in specifying which type of $c$ 
is actually a spacetime function: would it be the electromagnetic speed of 
light $c_{\rm{EM}} = 1/\sqrt{\epsilon_0 \mu_0}$ or the spacetime coupling $c_{\rm{ST}}$ 
appearing in the definition of the interval $ds^2$? Our proposal in this work aligns 
with the latter, i.e., $c=c_{\rm{ST}}$ throughout this paper. 
Other authors' criticisms on VSL scenarios weigh heavily on the argument that varying couplings should be actually 
dimensionless quantities---see, for example, Ref. \cite{Duff:2001ba}. The controversy did not prevent researchers of 
investigating the impact of varying couplings in cosmology. In fact, VSL proposals include the classical 
papers by Moffat \cite{Moffat:1992ud}, Albretch and Magueijo \cite{Albrecht:1998ir}, 
and also attempts to constrain an eventually
varying $c$ \cite{Salzano:2014lra,Cao:2016dgw,Salzano:2016hce,Qi:2014zja,Liu:2021eit,Liu:2018qrg,Mendonca:2021eux,Mukherjee:2023yxq}. 
On the other hand, varying-$G$ models and their constraints according to various data sets are explored in Refs.
\cite{Dirac:1937ti,Jofre:2006ug,Verbiest:2008gy,Lazaridis:2009kq,Garcia-Berro:2011kvq,Ooba:2016slp,Zhao:2018gwk,Vijaykumar:2020nzc}
(see Ref. \cite{Uzan:2002vq} for a review). 

One common feature of the VSL and varying-$G$ models cited above is that they do not 
account for simultaneous variations of $c$ and $G$: it is either $c=c(x^{\mu})$ or $G=G(x^{\mu})$, not both. 
Instead, Refs. \cite{Franzmann:2017nsc,Costa:2017abc,Gupta:2020anq,Gupta:2021wib,Gupta:2020wgz,Gupta:2021eyi,Gupta:2021tma,Gupta:2022amf,Bonanno:2020qfu,Lee:2020zts,Cuzinatto:2022vvy,Cuzinatto:2022mfe,Cuzinatto:2022dta}
 contemplate the possibility for two or all the couplings in the set $\{c,G,\Lambda\}$ varying 
at the same time. However, as we will see, such couplings do not  vary arbitrarily. 
In this sense, those couplings vary together, i.e. $\{c,G,\Lambda\}$ are \textit{co-varying physical couplings} (CPC). 
We will review the CPC framework as a modified gravity proposal in Section \ref{sec:CPC}. 
For now, let us motivate the subject of this paper. 

So far, we have mentioned cosmology and its challenges to advocate in favor of the fairness of 
considering modified gravity scenarios as alternatives to GR. Once one accepts that, it is necessary 
to investigate the impacts of modified gravity outside the realm of cosmology and into the scope of astrophysics. 
More specifically, GR predicted the existence of gravitational waves and black holes. 
 Each of such astrophysics predictions by 
GR were exquisitely confirmed by the LIGO-VIRGO collaboration with its detection of 
GW from binary inspirals \cite{LIGOScientific:2016aoc} and the EHT collaboration 
with its images of BHs shadows \cite{EventHorizonTelescope:2019dse,EventHorizonTelescope:2022wkp}.
This means that GR should be valid in the typical regimes of these events. 
Therefore, whatever is the modified gravity model candidate at hand, it should be tested against 
the data from these astrophysical sources. Either this candidate model is rejected by observations or 
constrained by the data. In this paper we sough to investigate the impact of CPC framework on black hole shadows.

The EHT collaboration published shadows of two black holes so far: 
M87* \cite{EventHorizonTelescope:2019dse}, the black hole in the 
center of the Messier 87 galaxy, and Sgr A* \cite{EventHorizonTelescope:2022wkp}, the black hole in the center
 of our Milky Way galaxy. These are supermassive rotating black holes with accretion discs of materials. 
According to EHT, both are described by Kerr solution \cite{Kerr:1963ud}.
 However, according to EHT collaboration, slowly rotating black holes, like Sgr A*, 
 can be approximated by the Schwarzschild line element \cite{Schwarzschild:1916uq}.\footnote{See 
 Ref. \cite{Perlick:2021aok} for a review on black hole shadows of static and non-static black holes.}
 
The shadow observation of distant black holes has to account for the cosmological redshift 
caused by the Hubble flow. In fact, as Bisnovatyi-Kogan and Tsupko \cite{Bisnovatyi-Kogan:2018vxl} 
pointed out in the context of GR, 
the black hole solution at cosmological distances should have its exterior solution (far away from the hole) 
complemented by the Friedmann-Lema\^itre-Robertson-Walker (FLRW) spacetime,
thus the shadow size should be affected by the cosmological expansion. 
In this regard, the McVittie black hole \cite{McVittie} was conceived to 
take in consideration the cosmological expansion. 
For values of the radial coordinate close to the black hole, the gravitational influence is strong, and 
the McVittie metric can turn into the Schwarzschild 
spacetime after suitable coordinate transformations; 
but, at large distances from the hole, the scale factor becomes an important ingredient 
of the black hole description. Thus, the McVittie metric is the appropriate metric for modeling a 
nonrotating or slowly rotating
black hole in an expanding universe and the shadow it casts. The shadow size measurement is performed via the 
angular diameter distance $d_A$; here the Universe expansion could also play an important role. 
The CPC framework modifies FLRW spacetime and, in principle, also affects the description 
of the Hubble flow \citep{Costa:2017abc,Gupta:2020anq}. For this reason, it is mandatory to 
generalize McVittie solution for accommodating the hypothesized co-varying couplings. 
Moreover, we shall have to adapt the angular diameter distance equation typical of standard cosmology 
to comply with the assumed varying couplings. Both changes (in the McVittie metric and in the $d_A$ formula) 
will add up to the new description of the black hole shadow within the context of CPC.
As we will see, the shadow angular radius could be either larger or smaller when compared to the shadow within
the GR context, where the couplings $\{c,G,\Lambda\}$ are constant.\footnote{Modifications on the angular
	size of black hole shadows are studied in alternative scenarios like the Lorentz symmetry breaking 
	model \citep{Maluf:2020kgf}, which is also a scenario of modified gravity.}

The remainder of the paper is organized as follows: Section \ref{sec:CPC} recalls the 
co-varying physical couplings framework wherein the set $\{c,G,\Lambda\}$ is assumed 
to vary concomitantly. This simultaneous variation is a consequence of the general 
constraint arising from the general covariance requirement upon our modified gravity 
model (Subsection \ref{subsec:GC}). Subsection \ref{subsec:Distances} addresses 
how cosmological distances are affected by the co-varying couplings in the CPC scenario. 
Section \ref{sec:McVittie} analyses the modification to McVittie black hole geometry due to the 
CPC modification of the field equations; it is shown how this solution constrains CPC scenario. 
The black hole shadows are finally characterized in Section \ref{sec:Shadow} where comparisons are made 
with the results expected from McVittie spacetime in GR. Our final comments are given in Section \ref{sec:Conclusion}.


\section{CPC framework} \label{sec:CPC}

We have mentioned in Introduction that astrophysical observations confirm the predictions of 
GR with astounding level of precision. Therefore, any modified gravity proposal should 
move away from GR as little as possible. In this spirit, we propose to keep the main assumptions of 
GR while relaxing the constancy of $\{c,G,\Lambda\}$.


\subsection{The general constraint} \label{subsec:GC}

The CPC framework is a modified gravity alternative wherein Einstein field equations hold, i.e., 
\begin{equation}
G_{\mu \nu} + \Lambda g_{\mu \nu} = \frac{8\pi G}{c^4} T_{\mu \nu}, 
\label{eq:EFE}
\end{equation}
while the Newtonian gravitational coupling $G$, the spacetime causality coupling $c$, and the cosmological parameter $\Lambda$ are not genuine constants but otherwise are allowed to vary with respect to the spacetime 
coordinates $x^{\mu}$, that is to say,  $G=G(x^{\mu})$, $c=c(x^{\mu})$, and  $\Lambda=\Lambda(x^{\mu})$. The Einstein tensor is unchanged, namely
\begin{equation}
G_{\mu \nu} = R_{\mu \nu} - \frac{1}{2} R g_{\mu \nu}, 
\label{eq:G_munu}
\end{equation}
so that Bianchi identity is kept:
\begin{equation}
\nabla_{\mu} G^{\mu \nu} = 0,
\label{eq:Bianchi}
\end{equation}
where $\nabla_{\mu}$ stands for the covariant derivative. 
By extracting the covariant derivative of Eq.~(\ref{eq:EFE}), using the metric compatibility 
condition $\nabla_{\mu} g^{\mu \nu} =0$, and assuming covariant conservation of the energy momentum tensor,
\begin{equation}
\nabla_{\mu} T^{\mu \nu} = 0, 
\label{Conserved_T}
\end{equation}
one has:
\begin{equation}
\left( \frac{\partial_\mu G}{G} - 4 \frac{\partial_\mu c}{c}  \right) \frac{8\pi G}{c^4} T^{\mu \nu} - \left( \partial_\mu \Lambda \right) g^{\mu \nu} = 0 .
\label{eq:GC}
\end{equation}
This relation is called the general constraint (GC). It entangles the eventual variations of $\{c,G,\Lambda\}$, 
thus the name CPC of our framework. Eq.~(\ref{eq:GC}) also couples the co-varying couplings 
to the matter-energy content through the product with $T^{\mu \nu}$ in the first term. In particular, 
in the presence of matter $T^{\mu \nu}\neq 0$ and if $\Lambda$ is a genuine constant 
$\left( \partial_\mu \Lambda \right) = 0$, it follows:
\begin{equation}
\left( \frac{\partial_\mu G}{G} - 4 \frac{\partial_\mu c}{c}  \right)  = 0 \Rightarrow \frac{G}{G_0} = \left( \frac{c}{c_0} \right)^4 \quad \left( \Lambda = \rm{const} \right).
\label{eq:minimal-CPC}
\end{equation}
Eq.~(\ref{eq:minimal-CPC}) characterizes the minimal CPC model, thoroughly studied in Ref.~\cite{Cuzinatto:2022mfe}. 
Inspired by this specific case (which had been already mentioned in Ref.~\cite{Costa:2017abc}), Gupta~\cite{Gupta:2020anq} suggests the following \textit{Ansatz} for the general case in which $T^{\mu\nu}\neq 0$ and $\partial_{\mu}\Lambda \neq 0$:
\begin{equation}
\left( \frac{\partial_\mu G}{G} - \sigma \frac{\partial_\mu c}{c}  \right)  = 0 \Rightarrow \frac{G}{G_0} = \left( \frac{c}{c_0} \right)^{\sigma} \quad \left( \sigma = \rm{const} \right).
\label{eq:Gupta-ansatz}
\end{equation}
Eqs.~(\ref{eq:EFE}), (\ref{eq:GC}) and (\ref{eq:Gupta-ansatz}) should be solved simultaneously for any given 
line element with open parameters. This will be performed for the McVittie geometry in Section \ref{sec:McVittie}. 
As we have said (Section \ref{sec:Intro}), the McVittie spacetime describes a black hole in a cosmological context. As we will see momentarily, the way the observer measures distances in such spacetime is impacted by the CPC framework. This is a particularly sensitive matter for the determination of the black hole angular size: the angular diameter distance $d_A$ is involved in its measurement. The next subsection builds the expression for $d_A$ appropriate for a CPC universe model.


\subsection{Distances in the CPC framework} \label{subsec:Distances}

Gupta ~\cite{Gupta:2020anq} has shown in detail two fundamental results regarding the contact of CPC framework with cosmological observations. Firstly, the way we calculate the redshift $z$ in the CPC framework is not different from standard cosmology \cite{Weinberg}:
\begin{equation}
z \equiv \frac{\lambda_0-\lambda_e}{\lambda_e} = \frac{a_0}{a_e} - 1.
\label{eq:z}
\end{equation}
Here $\lambda_e$ ($\lambda_0$) is the radiation wavelength at emission (observation) and $a_e = a(t_e)\equiv a(t)$ ($a_0 = a(t_0) \equiv 1$) is the value of the scale factor at emission (observation). Secondly, the expression for the proper distance within CPC cosmology reads:
\begin{equation}
d_p(z) = \frac{1}{H_0} \int^{z}_{0} \frac{c(z)}{E(z)} dz,
\label{eq:d_p}
\end{equation}
where $E(z) \equiv H(z)/H_0$ is the normalized Hubble parameter. The Hubble parameter 
is $H(t) = \frac{\dot{a}}{a}$, where the upper dot denotes differentiation with respect to the cosmic time $t$, 
and $H_0$ is its the present-day value.

Moreover, Cuzinatto et al. \cite{Cuzinatto:2022dta} devoted great case to build the relation between the luminosity distance $d_L$ and the proper distance  $d_p$ in the CPC context. The result is
\begin{equation}
d_L(z) = \left( \frac{c_e}{c_0} \right) (1+z) \, d_p(z),
\label{eq:d_L}
\end{equation}
where $c_e=c(z)$ is the speed of light at emission. Today, $c(0)=c_0$ because $t=t_0$ corresponds to $a(t_0)=a_0=1$ for $z=0$ (the present-day values). 
In standard cosmology, $c(z) = \rm{constant}$, thus $c_e = c_0$, and Eqs.~(\ref{eq:d_p}) and (\ref{eq:d_L}) reduce to the well-known results.

In this paper, we detail how the angular diameter distance $d_A$ is changed in the CPC context. The angular-diameter distance assumes the existence of a standard yardstick, an object analogous to the role played by standard candles with respect to luminosity and the luminosity distance $d_L$. 

A standard yardstick has a proper length $\ell$, its ends bearing comoving coordinates $\left(r,\theta_{1},\phi\right)$ and $\left(r,\theta_{2},\phi\right)$; it is subtended by an angle $\delta\theta$ by an observer at coordinates
$\left(0,0,0\right)$. The angular diameter distance $d_{A}$ is the distance from the observer to the yardstick as computed using the small-angle formula $\ell=d_{A}\delta\theta$, i.e.
\begin{equation}
d_{A}=\frac{\ell}{\delta\theta}.
\label{eq:d_A-definition}
\end{equation}
This is the proper distance in an Euclidian static universe. However, in an expanding universe, the proper 
length between the two ends of the yardstick should be given by the  FLRW line element \cite{Ryden},
\begin{equation}
ds^{2}=-c^{2}dt^{2}+a\left(t\right)^2\left(dr^{2}+S_{k}\left(r\right)^2 d\Omega^2 \right),
\label{eq:FLRW(S_k)}
\end{equation}
where $d\Omega^2=d\theta^2+\sin^2 \theta d\phi^2$ and
\begin{equation}
S_{k}\left(r\right)=\begin{cases}
R_0\sin\left(r/R_0\right)\,, & k=+1\\
r\,, & k=0\\
R_0\sinh\left(r/R_0\right)\,, & k=-1
\end{cases} \label{eq:S_k}
\end{equation}
with $dr=d\phi=0$. We recall that $k$ stands for the spacial curvature parameter (a flat spacetime means $k=0$).
 Also, the observation of the full length of the yardstick occurs at a single instant of time, meaning that $dt=0$. Notice that the speed of light that enters FLRW line element via the term $-c^{2}\left(t\right)dt^{2}$
does not contribute upon taking $dt=0$. On top of that, our small angle $\delta\theta$ should correspond to the angular differential $d\theta$ under the assumption $\ell=ds$, which is an invariant. Therefore,
\[
ds^{2}=a\left(t_{e}\right)^2\left[S_{k}\left(r\right)^2 d\theta^{2}\right]\qquad\left(\text{yardstick}\right),
\]
with $t=t_{e}$ the instant of time corresponding to the light emission by the standard yardstick (the proper time at the coordinates where $\ell=ds$ makes sense). Hence,
\begin{equation}
\ell=a\left(t_{e}\right)S_{k}\left(r\right)\delta\theta=\frac{S_{k}\left(r\right)\delta\theta}{\left(1+z\right)},\label{eq:proper-length}
\end{equation}
where the last equality follows from Eq.~(\ref{eq:z}). Then, combining Eqs.~(\ref{eq:d_A-definition}) and (\ref{eq:proper-length}), we have
\begin{equation}
d_{A}=\frac{S_{k}\left(r\right)}{\left(1+z\right)}.
\label{eq:d_A(S_k,z)}
\end{equation}
This is precisely what it is obtained for regular cosmology (without co-varying physical couplings). However, when one tries to substitute $S_{k}\left(r\right)$ in terms of the luminosity distance, the difference between the CPC framework and the standard result appears. In fact, Eq.~(\ref{eq:d_L}) is the same as
\begin{equation}
d_{L}=S_{k}\left(r\right)\left(1+z\right)\left(\frac{c_{e}}{c_{0}}\right),\label{eq:d_L(z,c)-coda}
\end{equation}
because $S_{k}(r) = r = a(t_0) r = d_p(t_0)$ in a flat space geometry ($k=0$), according to the reasoning in Ref. \cite{Ryden}.\footnote{For a curved space sector, $S_{k}\left(r\right)$ depends on the parameter $R_0$ as per Eq.~(\ref{eq:S_k}) wherein $R_0$ may be understood as the universe's radius of curvature. In a nearly flat universe, $R_0$ will be larger than the current horizon distance. Therefore, objects with finite redshift will be within the observable universe and have a proper distance smaller than the radius of curvature. In this case, we can make the assumption $r\ll R_0$ which allows us to approximate both $S_{k=+1}=R_0\sin\left(r/R_0\right)\approx R_0\left(r/R_0\right)=r$ and $S_{k=-1}=R_0\sinh\left(r/R_0\right)\approx R_0\left(r/R_0\right)=r$. This reasoning would suport $S_{k}\approx d_{p}\left(t_{0}\right)$ even in the cases $k=\pm1$.} Plugging (\ref{eq:d_L(z,c)-coda}) into (\ref{eq:d_A(S_k,z)}) yields:
\begin{equation}
d_{A}=\frac{d_{L}}{\left(1+z\right)^{2}}\left(\frac{c_{0}}{c_{e}}\right).\label{eq:d_A(d_L)}
\end{equation}
In regular cosmology $c_{e}=c_{0}$ and we recover the expected relation $d_{A}=d_{L}\left(1+z\right)^{-2}$. If the speed of light is greater in the past, $c_{e}>c_{0}$ yielding a reduced angular-diameter distance. The converse is true for a decreasing speed of light, which leads to an enhanced value of $d_{A}$.


\section{McVittie black hole in the CPC framework} \label{sec:McVittie}
The McVittie geometry \cite{McVittie} describes a black hole in an expanding universe.\footnote{See 
Nolan's articles \cite{Nolan:1998xs,Nolan:1999kk,Nolan:1999wf} 
on properties of the McVittie solution and a criticism in Ref. \cite{Kaloper:2010ec}. In particular,
there are studies on the casual structure \cite{daSilva:2012nh}, 
time-dependent mass parameter \cite{Maciel:2015dsh}, and charged black holes \cite{Guariento:2019ock} using
the McVittie geometry.}
The flat spacetime metric for this interesting solution of the Einstein 
field equations is given in the $(t,r,\theta,\phi)$
coordinates by
\begin{align}
ds^2= & -\left(\frac{1-\mu (t,r)}{1+\mu(t,r)} \right)^2 c^2dt^2+a(t)^2 \left(1+\mu(t,r) \right)^4 \nonumber \\
& \times \left(dr^2+r^2d\Omega^2 \right),
\label{McVittie}
\end{align}
where
\begin{equation}
\mu(t,r)=\frac{GM}{2c^2r a(t)},
\end{equation}
$a(t)$ is the scale factor, and $M$ is the black hole mass. 
The main idea in the McVittie spacetime is that for $r<r_l$ the black hole influence on spacetime should be
taken in consideration. On the other hand, for $r>r_l$ spacetime is asymptotically the flat FLRW spacetime. It is worth
pointing out that the limiting value of the radial coordinate is $r_l \gg GM/c^2$.

Let us assume here, in agreement with the CPC framework, that the physical couplings $c,\,G$ and
$\Lambda$ vary only with respect to time, that is,
\begin{equation}
c\rightarrow c(t), \hspace{0.5cm} G\rightarrow G(t), \hspace{0.25cm} \mbox{and} \hspace{0.25cm} \Lambda\rightarrow\Lambda(t).
\label{Parameters}
\end{equation}
Firstly, we adopt the following parametrization presented in Cuzinatto et al. \cite{Cuzinatto:2022dta}:
\begin{equation}
c(t)=c_0 \phi_c(a), \hspace{0.5cm} \Lambda(t)=\Lambda_0 \phi_{\Lambda}(a),
\label{CL}
\end{equation}
and Gupta's \textit{Ansatz} (\ref{eq:Gupta-ansatz}) for a time-dependent Newtonian parameter, i.e, 
\begin{equation}
G(t)=G_0\left(\frac{c(t)}{c_0} \right)^\sigma.
\label{G}
\end{equation}
The deviations $\phi_c$ and $\phi_{\Lambda}$ depend on the scale factor $a(t)$, and
 the subscript zero stands for the present-day values of the parameters. Therefore, $\phi_{c,0}=\phi_{\Lambda,0}=1$.
 Arguably, we have three new parameters to adjust, namely $\phi_c,\phi_{\Lambda}$ and $\sigma$. However, the parameter $\sigma$ is found out naturally using McVittie spacetime.
 
After solving the modified Einstein field equations (\ref{eq:EFE}) assuming the McVittie geometry and the
parameters (\ref{CL}) and (\ref{G}), 
there are two off-diagonal terms of the energy-momentum tensor, namely
\begin{equation}
T^{t}_{\ r}=g^{tt}T_{tr} \propto (\sigma - 2) \hspace{0.25cm} \mbox{and} \hspace{0.25cm} T^{r}_{\ t}=g^{rr}T_{rt} \propto (\sigma - 2),
\end{equation}
with $T^{t}_{\ r} \neq T^{r}_{\ t}$. In order to 
eliminate these off-diagonal terms, something absent in the general relativity context, 
we adopt Gupta's \textit{Ansatz} for $\sigma =2$. As we can read, 
contrary to the FLRW scenario \cite{Cuzinatto:2022dta}, 
the value of the parameter $\sigma$ appears regardless the data in the context studied here.
Therefore, throughout the remaining part of this article, we assume both $\sigma=2$ and the validity of the 
McVittie solution in the CPC context. Indeed, as we will see,  the McVittie spacetime satisfies both the 
the gravitational field equations (\ref{eq:EFE}) and the GC (\ref{eq:GC}).

With a diagonal energy-momentum, written in terms of the energy density $\varepsilon$ and pressure $p$, namely
 $T^{\mu}_{\ \nu}=\mbox{diag}(-\varepsilon,p,p,p)$, one has the energy density explicitly given by
\begin{equation}
\varepsilon(t)=\frac{c_0^2}{8\pi G_0}\left(3H^2-\Lambda_0 c_0^2 \phi_{\Lambda} \phi_{c}^2 \right).
\label{Density}
\end{equation}
This is one of Friedmann's equations for a flat spacetime in the CPC scenario studied 
in Cuzinatto et al. \cite{Cuzinatto:2022dta}.

As we said, McVittie spacetime should satisfy the GC (\ref{eq:GC}) in order to be a viable solution 
in the CPC framework. For parameters that vary only with respect to time, like those ones 
in Eqs. (\ref{CL})-(\ref{G}), the GC is simply
\begin{equation}
\left(\frac{\dot{G}}{G}-4\frac{\dot{c}}{c}\right)\frac{8\pi G}{c^4}T^{tt} - \dot{\Lambda}g^{tt}=0.
\label{GC2} 
\end{equation}
For the McVittie metric (\ref{McVittie}), the constraint (\ref{GC2}) reads
\begin{equation}
\left(\dot{a}^2- \frac{1}{3}c_0^2\Lambda_0 a^2\phi_{\Lambda}\phi_{c}^2 \right)\phi'_c - \frac{1}{6}c_0^2\Lambda_0a^2 \phi_{c}^3 \phi'_{\Lambda}=0,
\label{Constraint}
\end{equation}
where prime denotes derivative with respect to the scale factor (that is, $\phi'_{c}=d\phi_c/da$).
It is worth noting an important point regarding the GC given by Eq. (\ref{Constraint}): 
it is the same constraint from the FLRW metric in the CPC scenario \cite{Cuzinatto:2022dta}, thus it 
 carries just the cosmological side of the McVittie spacetime, that is to say, it does not
bring out the mass parameter $M$. As it happens, the varying couplings, cf. assumed in Eqs. (\ref{CL})--(\ref{G}),
are present only at large scales, for $r>r_l$, where the spacetime is asymptotically FLRW. Our calculations regarding the
shadow angular size will be made for those scales. 

Since the energy-momentum tensor is covariantly conserved in the CPC context, the continuity equation is still
true. For the McVittie spacetime, even in the CPC scenario, from the component $\nabla_{\mu} T^{0\mu}=0$, 
we have the continuity equation
\begin{equation}
\dot{\varepsilon}+3H\left(\frac{1-\mu (t,r)}{1+\mu(t,r)} \right)\left(\varepsilon+p\right)=0.
\label{Continuity}
\end{equation} 
As we are interested in cosmological distances, and the GC (\ref{Constraint}) excludes the 
black hole mass term, the metric 
function $\mu(t,r)$ is set to zero for $r>r_l$. Thus, Eq. (\ref{Continuity}) is the same continuity equation for the 
FLRW metric in both GR and the CPC context.  
As a consequence, a useful relation for rewriting the GC (\ref{Constraint}) comes from 
Eq. (\ref{Continuity})  for $\mu(t,r)=0$, namely
\begin{equation}
\varepsilon=\varepsilon_0 \left(\frac{a}{a_0} \right)^{-3(1+w)},
\label{varepsilon}
\end{equation}
in which the equation of state for the matter-energy content $p=w\varepsilon$ was adopted.

By assuming a universe made up of matter and
dark energy, that is, excluding the tiny contribution of radiation in the energy density, 
the GC (\ref{Constraint}), from Eq. (\ref{varepsilon}), could be better rewritten as
\begin{equation}
2\Omega_{m,0} a_0^{-1} a^{-3(1+w)} \phi_c^{-3}\phi'_{c} =\Omega_{\Lambda,0}\phi'_{\Lambda},
\label{Final_Constraint}
\end{equation}
where 
\begin{equation}
\Omega_{m,0}=\frac{\varepsilon_0}{\varepsilon_{c,0}}=1-\Omega_{\Lambda,0}\phi_{\Lambda,0}\phi_{c,0}^2=1-\Omega_{\Lambda,0}, 
\end{equation}
with
\begin{equation}
\varepsilon_{c,0}=\frac{3H_0^2c_0^2}{8\pi G_{0}} \hspace{0.25cm} \mbox{and} \hspace{0.25cm} \Omega_{\Lambda,0}=\frac{\Lambda_0 c_0^2}{3H_0^2}.
\end{equation}
As usual, $\Omega_{m,0}$ is the present-day matter density parameter, $\varepsilon_{c,0}$ is the
critical value of the energy density, and $\Omega_{\Lambda,0}$ is the density parameter of dark energy measured today. 
As expected for a flat metric, $\Omega_{m,0}+\Omega_{\Lambda,0}=1$. 

Modified theories of gravity aim to address some of the limitations of GR's cosmology mentioned in 
Introduction, including that regarding the description of the present-day acceleration. 
Accordingly, CPC scenario offers an alternative description of dark energy via a dynamical $\Lambda$ 
co-varying with $c$ and $G$. Thus, in order to solve (\ref{Final_Constraint}), 
we assume the same relation $\phi_{\Lambda}(a)$ adopted and justified in Ref. \cite{Cuzinatto:2022dta}, which is written as 
\begin{equation}
\phi_{\Lambda}=\phi_{\Lambda,0}+\phi_{\Lambda,1}\left(1-\frac{a}{a_0} \right)=1+\phi_{\Lambda,1}\left(\frac{z}{1+z} \right),
\label{phi_L}
\end{equation}
where the redshift relation $1+z=a_0/a(t)$ was used---cf. Eq.~(\ref{eq:z}). The main idea is to obtain (or fix) 
the value of  $\phi_{\Lambda,1}$ and then get $\phi_c$. We shall take $\vert \phi_{\Lambda,1}\vert \ll 1$ in accordance with the assumption that the cosmological term $\Lambda(t)$ deviates ever so slightly from its present-day value. Moreover, by setting $\phi_{\Lambda}=\phi_{\Lambda,0}=1$ for $z=0$, we recover $\Lambda(t)=\Lambda_0$ or the present-day value for that parameter.

Now we are interested in how the other parameters ($\phi_c$ and $G$) vary with the redshift $z$.
For a universe made up of  matter as dust $(w=0)$ and dark energy, 
Eq. (\ref{Final_Constraint}), with the aid of (\ref{phi_L}),
 presents the following solution for the deviation of the speed of light:
\begin{equation}
\phi_c (a) = \left[1-\frac{\left(1-\Omega_{m,0}  \right) a_0^4 \phi_{\Lambda,1}}{4\Omega_{m,0}} \left(1-\left(\frac{a}{a_0}\right)^4 \right) \right]^{-\frac{1}{2}}.
\end{equation}  
Again, by using the redshift relation, one could write $\phi_c$ as function of the redshift, namely
\begin{equation}
\phi_c (z) = \left[1-\frac{\left(1-\Omega_{m,0}  \right) a_0^4 \phi_{\Lambda,1}}{4\Omega_{m,0}} \left(1- \left( 1+z \right)^{-4} \right) \right]^{-\frac{1}{2}}.
\label{phi_c}
\end{equation}
Once again, for $z=0$, $\phi_{c}=\phi_{c,0}=1$, as expected today. 
Therefore, with $\phi_{\Lambda,1}$ assumed and $\phi_c(z)$ given by (\ref{phi_c}) calculated, 
the set $\{c,G,\Lambda \}$ varies simultaneously in agreement with CPC framework. In this regard,
the McVittie solution with time varying couplings is solution of the Einstein (modified) field equations and of the GC
in the CPC context.

From the parametrization adopted in this work, 
following Ref. \cite{Cuzinatto:2022dta}, the set of parameters to 
be constrained is indeed $\{H_0,\Omega_{m,0},\phi_{\Lambda,1},\sigma \}$ for a universe made up of 
matter and dark energy. However, Cuzinatto et al. \cite{Cuzinatto:2022dta} showed that by using
the $H(z)$ and SNe Ia-Pantheon data, the parameter $\sigma$ is not constrained in the cosmological
context. Then the authors of that paper made their analysis using $\sigma=3$, a value just suggested 
by previous references \cite{Cuzinatto:2022vvy,Gupta:2020anq,Gupta:2021wib,Gupta:2020wgz,Gupta:2021eyi,Gupta:2021tma,Eaves}, and consequently obtained $\phi_{\Lambda,1}=0.039^{+0.12}_{-0.095}$ at 95\% confidence level. 
In this regard, $\sigma=2$ is not ruled out by the recent data yet. As we saw, adding up a black hole in the cosmological
context provides a value of $\sigma$ without prior data or previous references.


\section{Black hole shadow in the CPC framework} \label{sec:Shadow}

In order to build the formula for the shadow angular radius of a nonrotating black hole in the CPC scenario, we begin
with the angular diameter distance (\ref{eq:d_A-definition}),
where $\ell$ is the proper diameter and $\Delta \theta$ is the angular diameter of the black hole shadow.
Assuming a flat solution ($k=0$) and the parametrization for the speed of light (\ref{CL}),
 Eq. (\ref{eq:d_A(d_L)}) yields---by using $d_p$ and $d_L$, given by Eqs. (\ref{eq:d_p}) and (\ref{eq:d_L}),
  respectively---the angular diameter distance in terms of the redshift $z$ 
 in the CPC framework adopted here, that is
\begin{equation}
d_A(z) = \frac{c_0}{a_0H_0 (1+z)}\int_{0}^{z}\frac{\phi_c(z) dz}{E(z)},
\label{dA_CPC}
\end{equation}
where the normalized Hubble parameter $E(z)$ reads
\begin{equation}
E(z)=\sqrt{\Omega_{m,0}(1+z)^3+(1-\Omega_{m,0})\phi_{\Lambda}(z)\phi_c(z)^2}.
\end{equation}
And, as we said, $w=0$ within a universe made up of matter and dark energy. 
With $\phi_c=\phi_{\Lambda}=1$, we recover
the $\Lambda$CDM results. 

We follow the Bisnovatyi-Kogan and Tsupko work \cite{Bisnovatyi-Kogan:2018vxl}, where
 the authors show numerically that the shadow angular diameter in the McVittie
spacetime is approximately equal to the shadow angular diameter of a Schwarzschild black hole immersed into an
expanding spacetime. Thus, the shadow proper radius or diameter adopted in this article will be the 
shadow proper radius of the Schwarzschild black hole. 

Calculated for the first time by Synge \cite{Synge}, the shadow proper radius of the Schwarzschild black 
hole---whether in GR or, as we will see, in the CPC context---is written as
\begin{equation}
r_{\text{sh}}=\frac{3\sqrt{3} \ G(t)M}{c(t)^2}=\frac{3\sqrt{3} \ G_0M}{c_0^2},
\label{Proper_radius}
\end{equation}
due to the Newtonian parameter ($\ref{G}$).
As we can see, $\sigma =2$ assures that $r_{\text{sh}}$ is invariant as well as the event horizon radius.
Therefore, the black hole physics in the CPC framework asks for $\sigma=2$.

\begin{figure}
\begin{center}
\includegraphics[trim=0.5cm 0.4cm 0.9cm 0cm, clip=true,scale=0.5]{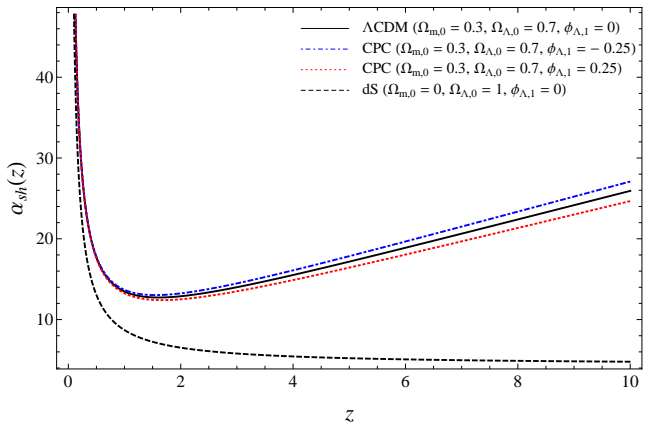}
\caption{The shadow angular radius of a Schwarzschild black hole in an expanding universe.
In this graphic, we compare the $\Lambda$CDM model with the CPC model.
As we can clearly see, in the CPC framework shadows could appear smaller or larger 
compared with the $\Lambda$CDM model. For a universe with matter ($\Omega_{m,0}>0$), both
models show that the angular radius is unbounded as the redshift increases. 
dS means de Sitter universe, a vacuum expanding and accelerating universe that is effect of the cosmological constant:
in this case, the angular size is bounded, but not zero as the redshift increases.
In this graphic, we adopted $c_0=G_0=a_0=H_0=M=1$.}
\label{Angular_Radius}
\end{center}
\end{figure}

Using (\ref{eq:d_A-definition}) and the proper radius (\ref{Proper_radius}), 
taking in consideration that $\ell = 2 r_{\text{sh}}$
and $\Delta \theta= 2 \alpha_{\text{sh}}$, 
we achieve the shadow
angular radius for a static black hole observed at cosmological distances ($r>r_l$) in the CPC scenario, namely
\begin{equation}
\alpha_{\text{sh}} (z) \approx \frac{r_{\text{sh}}}{d_A} = \frac{a_0H_0 r_{\text{sh}}(1+z)}{c_0} \left(\int_{0}^{z}\frac{\phi_c(z) dz}{E(z)} \right)^{-1}.
\label{Radius}
\end{equation}
As expected, for $\phi_c=\phi_{\Lambda}=1$, we have the same results obtained by Ref. \cite{Bisnovatyi-Kogan:2018vxl}
 in the $\Lambda$CDM model. Here, in the CPC framework,
the shadow of a spherically symmetric black hole could either increase or decreases 
with the redshift in comparison with
the angular size obtained from the $\Lambda$CDM model (see Fig. \ref{Angular_Radius}). This effect depends on
the value of $\phi_{\Lambda,1}$. If it is positive, the black hole shadow observed at cosmological distances
is smaller. On the other hand, if it is negative, the shadow is larger than the same shadow in the $\Lambda$CDM
cosmology.  

\begin{figure}
\begin{center}
\includegraphics[trim=0.4cm 0.4cm 0.65cm 0cm, clip=true,scale=0.57]{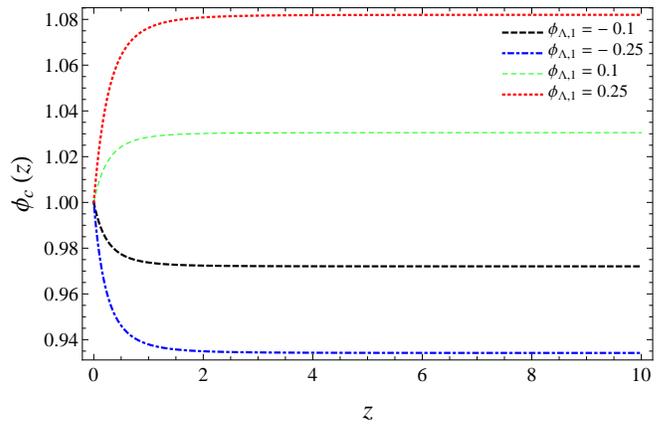}
\caption{The deviation of the speed of light, given by Eq. (\ref{phi_c}), for different values of the parameter
$\phi_{\Lambda,1}$ and redshift $z$. For negative values, the speed of light had its value increased during the cosmic
expansion. For positive values, one has the contrary situation.
In this graphic, we adopted $a_0=1$ and $\Omega_{m,0}=0.3$.}
\label{Phi}
\end{center}
\end{figure}

As for the value of $\phi_{\Lambda,1}$, a larger number of shadows captured by future EHT observations 
will provide at least an upper bound on that parameter. Slowly rotating
black holes like Sgr A*, observed at cosmological distances, will have their angular sizes described by Eq. (\ref{Radius}),
thus $\phi_{\Lambda,1}$, with the aid of $\phi_c(z)$, will be constrained. In this regard, this work is assumed
to be a theoretical model for future observations.   

Also, according to Fig. \ref{Angular_Radius}, we see the same behavior of the angular radius in both models
($\Lambda$CDM and CPC), as the 
redshift increases for a cosmological model with matter component $\Omega_{m,0}>0$. 
The angular size decreases as the redshift increases  until about $z=1.25$ and then starts to increase. 
This is a common behavior just as matter is present. For a single component universe,
in which matter is absent and the spacetime dynamics is driven by dark energy, the angular size 
decreases at a minimum value, but it is not zero. 

According to Eq. (\ref{eq:d_A(d_L)}), the angular diameter distance in the CPC framework 
is given in terms of emitted and present-day values of the 
speed of light, namely $d_A \propto c_0/c_e$. As Fig. (\ref{Phi}) shows and Eq. (\ref{phi_c}) indicates 
(keep in mind that $c(z)=c_0\phi_c(z)$), 
a positive value for the parameter $\phi_{\Lambda,1}$ means $c_e>c_0$, that is to say,
one has a smaller shadow angular radius and, at the same time, the speed of light decreased its
value during the universe expansion. For $\phi_{\Lambda,1}$ negative, the situation is opposite, because
the value of the speed of light increased during the cosmic expansion, that is $c_e<c_0$.  The very same conclusion
is suitable to the Newtonian parameter given by (\ref{G}). 
If $c_e<c_0$, it was smaller in the past; if $c_e>c_0$, it was larger
in the past.


\section{Final Comments} \label{sec:Conclusion}
In this work, the shadow angular radius of a static black hole was studied in the CPC framework, 
where the set of physical couplings given by the speed of light in vacuum, Newtonian parameter, and
the cosmological coupling $\{c,G,\Lambda \}$ could vary with the spacetime coordinates. 
Here we adopted a time variation of such parameters, thus $c(t)$, $G(t)$ and $\Lambda(t)$.
Consequently, we show that the McVittie spacetime is solution of the main constraint of the CPC framework.
Interestingly, in order to have the McVittie solution in the CPC context, the parameter $\sigma$---which relates
the Newtonian coupling and the speed of light in vacuum in recent CPC proposals---is 
naturally obtained  ($\sigma=2$), contrary to previous
works in which such a parameter is indirectly suggested. Indeed, as Ref. \cite{Cuzinatto:2022dta} states,
$\sigma=2$ is not ruled by the $H(z)$ and SNe Ia-Pantheon data.

By using a suitable parametrization in which small deviations for the physical couplings are assumed, 
we show that the shadow angular radius is very dependent on
the parameter $\phi_{\Lambda,1}$, which measures the cosmological coupling departure from a simple constant.
If $\phi_{\Lambda,1}$ is positive, the shadow angular radius is smaller when compared to the general relativity 
context or the $\Lambda$CDM model. If it is negative, one has the opposite situation, that is, the shadow angular
radius is larger. 
The same parameter $\phi_{\Lambda,1}$ could indicate
whether or not the speed of light in vacuum was larger or smaller in the past. If  $\phi_{\Lambda,1}>0$, 
$c(t)$ was larger in the past. On the other hand, if $\phi_{\Lambda,1}<0$, $c(t)$ was smaller. The same reasoning
is valid to the Newtonian coupling $G(t)$, which also depends on $\phi_{\Lambda,1}$.


\begin{acknowledgments}
RRC acknowledges the financial support from CNPq-Brazil (Grant 309984/2020-3). 
JCSN thanks the ICT-UNIFAL for the kind hospitality. 

\end{acknowledgments}



\begin{thebibliography}{10}

\bibitem{SupernovaSearchTeam:1998fmf}
A.~G.~Riess \textit{et al.} [Supernova Search Team],
Observational evidence from supernovae for an accelerating universe and a cosmological constant,
Astron. J. \textbf{116}, 1009 (1998). arXiv:astro-ph/9805201

\bibitem{SupernovaCosmologyProject:1998vns}
S.~Perlmutter \textit{et al.} [Supernova Cosmology Project],
Measurements of $\Omega$ and $\Lambda$ from 42 high redshift supernovae,
Astrophys. J. \textbf{517}, 565  (1999). arXiv:astro-ph/9812133 

\bibitem{Magana:2017nfs}
J.~Magana, M.~H.~Amante, M.~A.~Garcia-Aspeitia, and V.~Motta,
The Cardassian expansion revisited: constraints from updated Hubble parameter measurements and type Ia supernova data,
Mon. Not. Roy. Astron. Soc. \textbf{476}, 1036 (2018). arXiv:1706.09848

\bibitem{SDSS:2005xqv}
D.~J.~Eisenstein \textit{et al.} [SDSS],
Detection of the Baryon Acoustic Peak in the Large-Scale Correlation Function of SDSS Luminous Red Galaxies,
Astrophys. J. \textbf{633}, 560  (2005). arXiv:astro-ph/0501171 

\bibitem{SPT:2021vsu}
A.~B.~Mantz \textit{et al.} [SPT],
Cosmological constraints from gas mass fractions of massive, relaxed galaxy clusters,
Mon. Not. Roy. Astron. Soc. \textbf{510}, 131  (2021). arXiv:2111.09343

\bibitem{Liu:2022inf}
Y.~Liu \textit{et al.},
Gamma-Ray Burst Constraints on Cosmological Models from the Improved Amati Correlation,
Astrophys. J. \textbf{935}, 7 (2022). arXiv:2207.00455

\bibitem{Lusso:2020pdb}
E.~Lusso \textit{et al.},
Quasars as standard candles III. Validation of a new sample for cosmological studies,
Astron. Astrophys. \textbf{642}, A150  (2020). arXiv:2008.08586


\bibitem{Planck:2018vyg}
N.~Aghanim \textit{et al.} [Planck],
Planck 2018 results. VI. Cosmological parameters,
Astron. Astrophys. \textbf{641}, A6 (2020). arXiv:1807.06209

\bibitem{DiValentino:2021izs}
E.~Di Valentino, O.~Mena, S.~Pan, L.~Visinelli, W.~Yang, A.~Melchiorri, D.~F.~Mota, A.~G.~Riess and J.~Silk,
In the realm of the Hubble tension\textemdash{}a review of solutions,
Class. Quant. Grav. \textbf{38}, 153001 (2021). arXiv:2103.01183

\bibitem{Abdalla:2022yfr}
E.~Abdalla, G.~Franco Abell\'an, A.~Aboubrahim, A.~Agnello, O.~Akarsu, Y.~Akrami, G.~Alestas, D.~Aloni, L.~Amendola and L.~A.~Anchordoqui, \textit{et al.}
Cosmology intertwined: A review of the particle physics, astrophysics, and cosmology associated with the cosmological tensions and anomalies, 
JHEAp \textbf{34}, 49 (2022). arXiv:2203.06142 

\bibitem{Klypin:1999uc}
A.~A.~Klypin, A.~V.~Kravtsov, O.~Valenzuela and F.~Prada,
Where are the missing Galactic satellites?,
Astrophys. J. \textbf{522}, 82 (1999). arXiv:astro-ph/9901240

\bibitem{Carroll:2000fy}
S.~M.~Carroll,
The Cosmological constant,
Living Rev. Rel. \textbf{4}, 1 (2001).  arXiv:astro-ph/0004075

\bibitem{Coc:2017pxv}
A.~Coc and E.~Vangioni,
Primordial nucleosynthesis,
Int. J. Mod. Phys. E \textbf{26}, 1741002 (2017). arXiv:1707.01004 

\bibitem{Starobinsky:1980te}
A.~A.~Starobinsky,
A New Type of Isotropic Cosmological Models Without Singularity,
Phys. Lett. B \textbf{91}, 99 (1980).

\bibitem{Guth:1980zm}
A.~H.~Guth,
The Inflationary Universe: A Possible Solution to the Horizon and Flatness Problems,
Phys. Rev. D \textbf{23}, 347  (1981).

\bibitem{Linde:1981mu}
A.~D.~Linde,
A New Inflationary Universe Scenario: A Possible Solution of the Horizon, Flatness, Homogeneity, Isotropy and Primordial Monopole Problems,
Phys. Lett. B \textbf{108}, 389 (1982).

\bibitem{Mukhanov:1981xt}
V.~F.~Mukhanov and G.~V.~Chibisov,
Quantum Fluctuations and a Nonsingular Universe,
JETP Lett. \textbf{33}, 532  (1981).

\bibitem{Lehners:2008vx}
J.~L.~Lehners,
Ekpyrotic and Cyclic Cosmology,
Phys. Rept. \textbf{465}, 223  (2008). arXiv:0806.1245 

\bibitem{Planck:2018jri}
Y.~Akrami \textit{et al.} [Planck],
Planck 2018 results. X. Constraints on inflation,
Astron. Astrophys. \textbf{641}, A10 (2020). arXiv:1807.06211 

\bibitem{Kaluza:1921tu}
T.~Kaluza,
Zum Unit\"atsproblem der Physik,
Sitzungsber. Preuss. Akad. Wiss. Berlin (Math. Phys. ) \textbf{1921}, 966  (1921). \url{arXiv:1803.08616}

\bibitem{Klein:1926tv}
O.~Klein,
Quantum Theory and Five-Dimensional Theory of Relativity. (In German and English),
Z. Phys. \textbf{37}, 895 (1926).

\bibitem{CANTATA:2021ktz}
E.~N.~Saridakis \textit{et al.} [CANTATA],
Modified Gravity and Cosmology: An Update by the CANTATA Network,
Springer, 2021. arXiv:2105.12582

\bibitem{Minkevich:2016wya}
A.~V.~Minkevich,
Relationship of gauge gravitation theory in Riemann-Cartan space-time and general relativity,
Grav. Cosmol. \textbf{23}, 311 (2017). arXiv:1609.05285 

\bibitem{Aldrovandi:2013wha}
R.~Aldrovandi and J.~G.~Pereira,
\textit{Teleparallel Gravity: An Introduction} (Springer, Dordrecht, 2013).


\bibitem{Cuzinatto:2021ttc}
R.~R.~Cuzinatto, E.~M.~De Morais, and B.~M.~Pimentel,
Lyra scalar-tensor theory: A scalar-tensor theory of gravity on Lyra manifold,
Phys. Rev. D \textbf{103}, 124002 (2021). arXiv:2104.06295

\bibitem{Clifton:2011jh}
T.~Clifton, P.~G.~Ferreira, A.~Padilla and C.~Skordis,
Modified Gravity and Cosmology,
Phys. Rept. \textbf{513}, 1 (2012). arXiv:1106.2476 

\bibitem{Sotiriou:2008rp}
T.~P.~Sotiriou and V.~Faraoni,
f(R) Theories Of Gravity,
Rev. Mod. Phys. \textbf{82}, 451 (2010). arXiv:0805.1726

\bibitem{Nojiri:2010wj}
S.~Nojiri and S.~D.~Odintsov,
Unified cosmic history in modified gravity: from F(R) theory to Lorentz non-invariant models,
Phys. Rept. \textbf{505}, 59 (2011).  \\ arXiv:1011.0544

\bibitem{Capozziello:2011et}
S.~Capozziello and M.~De Laurentis,
Extended Theories of Gravity,
Phys. Rept. {509}, 167 (2011). arXiv:1108.6266

\bibitem{Buchbinder:2021wzv}
I.~L.~Buchbinder and I.~Shapiro,
\textit{Introduction to Quantum Field Theory with Applications to Quantum Gravity},
(Oxford University Press, Oxford, 2023.)

\bibitem{Cuzinatto:2016ehv}
R.~R.~Cuzinatto, C.~A.~M.~de Melo, L.~G.~Medeiros and P.~J.~Pompeia,
Scalar-multi-tensorial equivalence for higher order  $f\left( R,\nabla_{\mu} R,\nabla_{\mu_{1}}\nabla_{\mu_{2}}R,...,\nabla_{\mu_{1}}...\nabla_{\mu_{n} }R\right)$ theories of gravity,
Phys. Rev. D \textbf{93}, 124034 (2016). arXiv:1603.01563

\bibitem{Cuzinatto:2018chu}
R.~R.~Cuzinatto, C.~A.~M.~de Melo, L.~G.~Medeiros and P.~J.~Pompeia,
$f\left(R,\nabla_{\mu_{1}}R,...,\nabla_{\mu_{1}}...\nabla_{\mu_{n}}R\right)$ theories of gravity in Einstein frame: a higher order modified Starobinsky inflation model in the Palatini approach,
Phys. Rev. D \textbf{99}, 084053 (2019). arXiv:1806.08850 

\bibitem{Faraoni:2004pi}
V.~Faraoni, 
\textit{Cosmology in Scalar-Tensor Gravity},
(Kluwer Academic Publishers, Dordrecht, 2004).

\bibitem{Brans:1961sx}
C.~Brans and R.~H.~Dicke,
Mach's principle and a relativistic theory of gravitation,
Phys. Rev. \textbf{124}, 925 (1961).

\bibitem{Barbour:2010dp}
J.~Barbour,
The Definition of Mach's Principle,
Found. Phys. \textbf{40}, 1263 (2010). arXiv:1007.3368 

\bibitem{Caldwell:1997ii}
R.~R.~Caldwell, R.~Dave and P.~J.~Steinhardt,
Cosmological imprint of an energy component with general equation of state,
Phys. Rev. Lett. \textbf{80}, 1582 (1998). arXiv:astro-ph/9708069 

\bibitem{Ellis:2003pw}
G.~F.~R.~Ellis and J.~P.~Uzan,
`c' is the speed of light, isn't it?,
Am. J. Phys. \textbf{73}, 240 (2005). arXiv:gr-qc/0305099

\bibitem{Duff:2001ba}
M.~J.~Duff, L.~B.~Okun and G.~Veneziano,
Trialogue on the number of fundamental constants,
JHEP \textbf{03}, 023 (2002). arXiv:physics/0110060 

\bibitem{Moffat:1992ud}
J.~W.~Moffat,
Superluminary universe: A Possible solution to the initial value problem in cosmology,
Int. J. Mod. Phys. D \textbf{2}, 351 (1993).  arXiv:gr-qc/9211020

\bibitem{Albrecht:1998ir}
A.~Albrecht and J.~Magueijo,
A Time varying speed of light as a solution to cosmological puzzles,
Phys. Rev. D \textbf{59}, 043516 (1999).  arXiv:astro-ph/9811018 

\bibitem{Salzano:2014lra}
V.~Salzano, M.~P.~Dabrowski and R.~Lazkoz,
Measuring the speed of light with Baryon Acoustic Oscillations,
Phys. Rev. Lett. \textbf{114}, 101304 (2015). arXiv:1412.5653 

\bibitem{Cao:2016dgw}
S.~Cao, M.~Biesiada, J.~Jackson, X.~Zheng, Y.~Zhao and Z.~H.~Zhu,
Measuring the speed of light with ultra-compact radio quasars,
JCAP \textbf{02}, 012 (2017). arXiv:1609.08748

\bibitem{Salzano:2016hce}
V.~Salzano,
Recovering a redshift-extended varying speed of light signal from galaxy surveys,
Phys. Rev. D \textbf{95}, 084035 (2017). arXiv:1604.03398 

\bibitem{Qi:2014zja}
J.~Z.~Qi, M.~J.~Zhang and W.~B.~Liu,
Observational constraint on the varying speed of light theory,
Phys. Rev. D \textbf{90}, 063526 (2014). arXiv:1407.1265

\bibitem{Liu:2021eit}
T.~Liu, S.~Cao, M.~Biesiada, Y.~Liu, Y.~Lian and Y.~Zhang,
Consistency testing for invariance of the speed of light at different redshifts: the newest results from strong lensing and Type Ia supernovae observations,
Mon. Not. Roy. Astron. Soc. \textbf{506}, 2181 (2021). arXiv:2106.15145

\bibitem{Liu:2018qrg}
Y.~Liu and B.~Q.~Ma,
Light speed variation from gamma ray bursts: criteria for low energy photons,
Eur. Phys. J. C \textbf{78}, 825 (2018). arXiv:1810.00636

\bibitem{Mendonca:2021eux}
I.~E.~C.~R.~Mendon\c{c}a, K.~Bora, R.~F.~L.~Holanda, S.~Desai and S.~H.~Pereira,
A search for the variation of speed of light using galaxy cluster gas mass fraction measurements,
JCAP \textbf{11}, 034 (2021).  arXiv:2109.14512 

\bibitem{Mukherjee:2023yxq}
P.~Mukherjee, G.~Rodrigues and C.~Bengaly,
Do current cosmological observations hint at the speed of light variability?, arXiv:2302.00867 

\bibitem{Uzan:2002vq}
J.~P.~Uzan,
The Fundamental Constants and Their Variation: Observational Status and Theoretical Motivations,
Rev. Mod. Phys. \textbf{75}, 403 (2003).  arXiv:hep-ph/0205340 

\bibitem{Dirac:1937ti}
P.~A.~M.~Dirac,
The Cosmological constants,
Nature \textbf{139}, 323 (1937).

\bibitem{Jofre:2006ug}
P.~Jofre, A.~Reisenegger and R.~Fernandez,
Constraining a possible time-variation of the gravitational constant through gravitochemical heating of neutron stars,
Phys. Rev. Lett. \textbf{97}, 131102 (2006). arXiv:astro-ph/0606708 

\bibitem{Verbiest:2008gy}
J.P.W.~Verbiest \textit{et al.}  
Precision timing of PSR J0437-4715: an accurate pulsar distance, a high pulsar mass and a limit on the variation of Newton's gravitational constant,
Astrophys. J. \textbf{679}, 675 (2008). arXiv:0801.2589 

\bibitem{Lazaridis:2009kq}
K.~Lazaridis \textit{et al.}
Generic tests of the existence of the gravitational dipole radiation and the variation of the gravitational constant,
Mon. Not. R. Astron. Soc. \textbf{400}, 805 (2009). arXiv:0908.0285 

\bibitem{Garcia-Berro:2011kvq}
E.~Garcia-Berro, P.~Loren-Aguilar, S.~Torres, L.~G.~Althaus and J.~Isern,
An upper limit to the secular variation of the gravitational constant from white dwarf stars,
JCAP \textbf{05}, 021 (2011). arXiv:1105.1992 

\bibitem{Ooba:2016slp}
J.~Ooba, K.~Ichiki, T.~Chiba and N.~Sugiyama,
Planck constraints on scalar-tensor cosmology and the variation of the gravitational constant,
Phys. Rev. D \textbf{93}, 122002 (2016). arXiv:1602.00809 

\bibitem{Zhao:2018gwk}
W.~Zhao, B.~S.~Wright and B.~Li,
Constraining the time variation of Newton's constant $G$ with gravitational-wave standard sirens and supernovae,
JCAP \textbf{10}, 052 (2018). arXiv:1804.03066

\bibitem{Vijaykumar:2020nzc}
A.~Vijaykumar, S.~J.~Kapadia and P.~Ajith,
Constraints on the time variation of the gravitational constant using gravitational-wave observations of binary neutron stars,
Phys. Rev. Lett. \textbf{126}, 141104 (2021). arXiv:2003.12832

\bibitem{Franzmann:2017nsc}
G.~Franzmann,
Varying fundamental constants: a full covariant approach and cosmological applications, \\
arXiv:1704.07368.

\bibitem{Costa:2017abc}
R.~Costa, R.~R.~Cuzinatto, E.~M.~G.~Ferreira and G.~Franzmann,
Covariant c-flation: a variational approach,
Int. J. Mod. Phys. D \textbf{28}, 1950119 (2019). arXiv:1705.03461 

\bibitem{Gupta:2020anq}
R.~P.~Gupta,
Cosmology with relativistically varying physical constants,
Mon. Not. Roy. Astron. Soc. \textbf{498}, 4481 (2020). arXiv:2009.08878

\bibitem{Gupta:2021wib}
R.~P.~Gupta,
Testing the Speed of Light Variation with Strong Gravitational Lensing of SNe 1a,
Res. Notes AAS \textbf{5},  176 (2021).

\bibitem{Gupta:2020wgz}
R.~P.~Gupta,
Varying physical constants and the lithium problem,
Astropart. Phys. \textbf{129}, 102578 (2021). \\ arXiv:2010.13628 

\bibitem{Gupta:2021eyi}
R.~P.~Gupta,
Faint young Sun problem and variable physical constants,
Mon. Not. Roy. Astron. Soc. \textbf{509}, 4285  (2021).

\bibitem{Gupta:2021tma}
R.~P.~Gupta,
Effect of evolving physical constants on type Ia supernova luminosity,
Mon. Not. Roy. Astron. Soc. \textbf{511}, 4238 (2022). arXiv:2112.10654

\bibitem{Gupta:2022amf}
R.~P.~Gupta,
Constraining variability of coupling constants with bright and extreme quasars,
Mon. Not. Roy. Astron. Soc. \textbf{513}, 5559 (2022). arXiv:2202.12758 

\bibitem{Bonanno:2020qfu}
A.~Bonanno, G.~Kofinas and V.~Zarikas,
Effective field equations and scale-dependent couplings in gravity,
Phys. Rev. D \textbf{103}, 104025 (2021). arXiv:2012.05338 

\bibitem{Lee:2020zts}
S.~Lee,
The minimally extended Varying Speed of Light (meVSL),
JCAP \textbf{08}, 054 (2021). arXiv:2011.09274 

\bibitem{Cuzinatto:2022dta}
R.~R.~Cuzinatto, R.~P.~Gupta, R.~F.~L.~Holanda, J.~F.~Jesus and S.~H.~Pereira,
Testing a varying-\ensuremath{\Lambda} model for dark energy within co-varying physical couplings framework,
Mon. Not. Roy. Astron. Soc. \textbf{515}, no.4, 5981 (2022). arXiv:2204.10764

\bibitem{Cuzinatto:2022mfe}
R.~R.~Cuzinatto, R.~F.~L.~Holanda and S.~H.~Pereira,
Observational constraints on varying fundamental constants in a minimal CPC model,
Mon. Not. Roy. Astron. Soc. \textbf{519}, 633 (2023). arXiv:2202.01371

\bibitem{Cuzinatto:2022vvy}
R.~R.~Cuzinatto, R.~P.~Gupta and P.~J.~Pompeia,
Dynamical Analysis of the Covarying Coupling Constants in Scalar\textendash{}Tensor Gravity,
Symmetry \textbf{15}, 709 (2023). arXiv:2204.00119

\bibitem{LIGOScientific:2016aoc}
B.~P.~Abbott \textit{et al.} [LIGO Scientific and Virgo],
Observation of Gravitational Waves from a Binary Black Hole Merger,
Phys. Rev. Lett. \textbf{116}, 061102 (2016).  arXiv:1602.03837 

\bibitem{EventHorizonTelescope:2019dse}
K.~Akiyama \textit{et al.} [Event Horizon Telescope],
First M87 Event Horizon Telescope Results. I. The Shadow of the Supermassive Black Hole,
Astrophys. J. Lett. \textbf{875}, L1 (2019). arXiv:1906.11238 

\bibitem{EventHorizonTelescope:2022wkp}
K.~Akiyama \textit{et al.} [Event Horizon Telescope],
First Sagittarius A* Event Horizon Telescope Results. I. The Shadow of the Supermassive Black Hole in the Center of the Milky Way,
Astrophys. J. Lett. \textbf{930}, no.2, L12 (2022).

\bibitem{Kerr:1963ud}
R.~P.~Kerr,
Gravitational field of a spinning mass as an example of algebraically special metrics,
Phys. Rev. Lett. \textbf{11}, 237 (1963).

\bibitem{Schwarzschild:1916uq}
K.~Schwarzschild,
Über das Gravitationsfeld eines Mass-\\enpunktes nach der Einsteinschen Theorie,
Sitzungsber. Preuss. Akad. Wiss. Berlin  \textbf{1916}, 189-196 (1916). arXiv:physics/9905030 

\bibitem{Perlick:2021aok}
V.~Perlick and O.~Y.~Tsupko,
Calculating black hole shadows: Review of analytical studies,
Phys. Rept. \textbf{947}, 1-39 (2022). arXiv:2105.07101

\bibitem{Bisnovatyi-Kogan:2018vxl}
G.~S.~Bisnovatyi-Kogan and O.~Y.~Tsupko,
Shadow of a black hole at cosmological distances,
Phys. Rev. D \textbf{98}, 084020 (2018). arXiv:1805.03311 

\bibitem{McVittie}
G.~C.~McVittie,
The mass-particle in an expanding universe,
Mon. Not. Roy. Astron. Soc. \textbf{93}, 325-339 (1933).

\bibitem{Maluf:2020kgf}
R.~V.~Maluf and J.~C.~S.~Neves,
Black holes with a cosmological constant in bumblebee gravity,
Phys. Rev. D \textbf{103}, no.4, 044002 (2021). arXiv:2011.12841 

\bibitem{Weinberg}
S. Weinberg, \textit{Cosmology} (Oxford University Press, New York, 2014).

\bibitem{Ryden}
B. Ryden, \textit{Introduction to Cosmology} (Cambridge University Press, Cambridge, 2017).

\bibitem{Nolan:1998xs}
B.~C.~Nolan,
A Point mass in an isotropic universe: Existence, uniqueness and basic properties,
Phys. Rev. D \textbf{58}, 064006 (1998). arXiv:gr-qc/9805041

\bibitem{Nolan:1999kk}
B.~C.~Nolan,
A Point mass in an isotropic universe. 2. Global properties,
Class. Quant. Grav. \textbf{16}, 1227-1254 (1999).

\bibitem{Nolan:1999wf}
B.~C.~Nolan,
A Point mass in an isotropic universe. 3. The region R less than or = to 2m,
Class. Quant. Grav. \textbf{16}, 3183-3191 (1999). arXiv:gr-qc/9907018 

\bibitem{Kaloper:2010ec}
N.~Kaloper, M.~Kleban and D.~Martin,
McVittie's Legacy: Black Holes in an Expanding Universe,
Phys. Rev. D \textbf{81}, 104044 (2010). arXiv:1003.4777

\bibitem{daSilva:2012nh}
A.~M.~da Silva, M.~Fontanini and D.~C.~Guariento,
How the expansion of the universe determines the causal structure of McVittie spacetimes,
Phys. Rev. D \textbf{87}, no.6, 064030 (2013). arXiv:1212.0155 

\bibitem{Maciel:2015dsh}
A.~Maciel, D.~C.~Guariento and C.~Molina,
Cosmological black holes and white holes with time-dependent mass,
Phys. Rev. D \textbf{91}, no.8, 084043 (2015). arXiv:1502.01003 

\bibitem{Guariento:2019ock}
D.~C.~Guariento, A.~Maciel, M.~M.~C.~Mello and V.~T.~Zanchin,
Charged cosmological black holes: a thorough study of a family of solutions,
Phys. Rev. D \textbf{100}, no.10, 104050 (2019). arXiv:1908.04961 

\bibitem{Eaves}
R. E. Eaves, 
Constraints on variation in the speed of light based on gravitational constant constraints, 
Mon. Not. Roy. Astron. Soc. \textbf{505}, 3590 (2021).

\bibitem{Synge}
J.~L.~Synge,
The Escape of Photons from Gravitationally Intense Stars,
Mon. Not. Roy. Astron. Soc. \textbf{131}, 463 (1966).


\end{thebibliography}
\end{document}